\documentclass[a4paper,twocolumn]{esapub}

\usepackage{natbib}
\usepackage{graphics}
\usepackage{amssymb} 
\usepackage[mathcal]{eucal}

\newcommand{\apj} {ApJ}
\newcommand{\mnras} {MNRAS}
\newcommand{\aap} {A\&A}

\newcommand{\aaps} {A\&AS}

\newcommand{\eqn} [1] {
\begin{equation} 
#1 
\end{equation}}

\newcommand{\eqna} [1] {
\begin{eqnarray} 
#1 
\end{eqnarray}}
\newcommand{\derivp} [2] {\frac {\partial #1 } {\partial #2} }

\title{Oscillation power as a test of stellar turbulence : Scanning the HR diagram}
\author{R\'eza Samadi}
\author{M.-J. Goupil}
\author{Y. Lebreton}
\author{A. Baglin}
\affil{Observatoire de Paris-Meudon, 5 place Jules Janssen, F-92195 Meudon,France}

\begin{document}

\maketitle

\begin{abstract}
The acoustic power injected by turbulent convection into solar-like
oscillations depends on the details of the turbulent spectrum. 
A  theoretical formulation for the oscillation power 
is developed which  generalizes previous ones.
The formulation  is first calibrated on a solar model in 
such a way as to reproduce the solar seismic data.
This allows to investigate different assumptions about the stellar turbulent spectrum. 
We next explore consequences of the assumed turbulent description for
some potentially solar-like oscillating stars. 
Large differences are found in the oscillation power of a given star 
when using different  turbulent spectra as well as 
in a  star to star comparison. Space seismic observations of such stars
will be valuable for discriminating between several turbulent models. 

\end{abstract}

\section{Introduction}

The acoustic power injected into the $p$ modes by turbulent convection
 has been modeled by several authors \citep{Balmforth92c,GMK94,GK77}.
In \citet{Samadi00I} a general formulation is proposed
which allows  to investigate more consistently 
different assumptions about the stellar turbulence 
such as its turbulent energy spectrum.

Providing that accurate measurements of the oscillation amplitudes and
 damping rates are available it is possible to evaluate the power
 injected into the modes and thus- by comparison
with the observations- to constrain current theories \citep{Samadi00}.

In the present paper, the formulation, viewed as a diagnostic tool of 
stellar turbulent spectra, 
is used to compute oscillations power of several solar-like stars.
We show  that  the expected low detection threshold 
of a space seismic experiment such as COROT \citep{Baglin98} for instance will provide
highly accurate  acoustic power spectra and 
therefore important information  on the associated turbulent
spectrum if  the   solar-like oscillating targets  
are  properly chosen.

\section{Power injected into  solar-like oscillations}

\subsection{Stochastic excitation}

The acoustic power injected into the oscillations is defined 
\citep[e.g.][]{GMK94} in terms of the damping rate $\eta$, the mean-square amplitude 
$\langle A^2 \rangle$, the mode inertia $I$ and oscillation frequency 
$\omega$ : 
\eqna{
P(\omega) = \eta\;  {\langle A^2 \rangle}\;I\;\omega^2
\label{eq:P_omega_0}
}
The mean-square amplitude accounts for both the excitation by turbulent convection and the damping processes. It can be written in a simplified expression as
\eqn{
\left < A^2 \right > \propto\eta^{-1}\int_{0}^{M}dm \; \rho_0 \; w^4 \;\left
(\derivp { \xi_r} {r} \right )^2  \;   \mathcal{S}(\omega,m)
\label{eqn:A2}
} 
where $\displaystyle{\xi_r }$ is the radial displacement eigenfunction, $\rho_0$ the density, $w$  the vertical rms velocity of the convective elements and $\mathcal{S}$ the turbulent  source function accounting for both the Reynolds and the entropy fluctuations.
Detailed expressions for $\left < A^2 \right >$ and $\mathcal{S}$ are given in 
\citet{Samadi00I}. The source function 
involves the turbulent kinetic energy spectrum $E(k)$ and the 
turbulent spectrum $E_s(k)$ of the entropy fluctuations 
which  can be related to $E(k)$  \citep[e.g.][]{Samadi00II}. The turbulent 
spectra in $\mathcal{S}$ are integrated over all eddy wavenumbers $k$  
and $\mathcal{S}$ is in turn integrated in Eq.(\ref{eqn:A2}) over the stellar mass $M$.

\subsection{Models for the solar turbulence}
\label{sec:Models for the solar turbulence}
 From the turbulence theory it is expected  that $E(k)$ 
follows the Kolmogorov spectrum as $E(k) \propto k^{p}$ 
with the slope $p=-5/3$. Observations of the solar granulation 
allow one to determine the turbulent kinetic spectrum $E(k)$ of the Sun.
Observations of the solar granulation by 
\citet{Espagnet93} and \citet{Nesis93}.
 confirm the existence of a turbulent cascade with $p \simeq -5/3$. 
On the other hand,  in the $k$ range (small $k$) prior to the turbulent cascade of slope  $p \simeq
-5/3$,  \citet{Espagnet93} determined a regime with a slope close to $0.7$ 
whereas \citet{Nesis93} found a slope of $p=-5$.
    \begin{figure}[ht]
	\resizebox{\hsize}{!}{\includegraphics{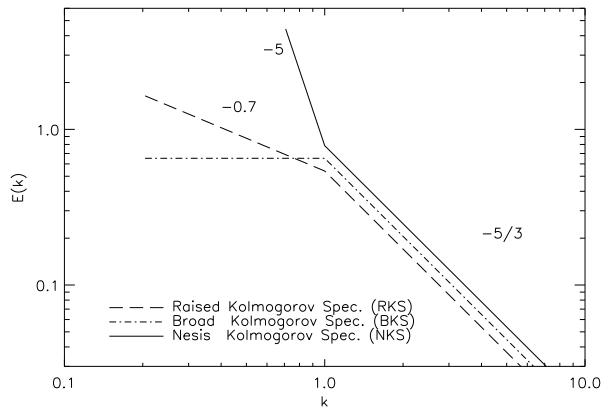}}
	\caption{Kinetic turbulent spectra versus wavenumber $k$.}
	\label{fig:spc_cinetique}
        \end{figure}

The spectrum observed by \citet{Espagnet93} is modeled  by
the so called ``Raised Kolmogorov Spectrum `` (RKS hereafter) 
as suggested by \citet{Musielak94}.
We also consider two additional spectra:  
the ``Nesis Kolmogorov Spectrum'' (NKS hereafter) - 
in agreement with the observations of \citet{Nesis93} - 
and the ``Broad Kolmogorov Spectrum'' (BKS hereafter).  
These spectra (Figure \ref{fig:spc_cinetique}) obey the Kolmogorov law 
for $k \geq k_0$ where $k_0$ is the wavenumber at which the turbulent cascade begins.

The wavenumber $k_0$ is unknown but can be related  to the mixing length
 $\Lambda$ as $k_0=2\pi \, / \, (\beta \Lambda)$ where  $\beta$ is  a 
free parameter introduced for the arbitrariness of such definition \citep{Samadi00I}.
Moreover the definition of the eddy time correlation, which
 enters the description of the
 turbulent excitation,  is somewhat  
arbitrary and is therefore  gauged by introducing an additional free parameter
 $\lambda$. 
We show that the oscillation power computed on the Sun is very sensitive to the values of the free parameters $\lambda$,  $\beta$  in \citet{Samadi00II}. 

\subsection{The solar case: comparison with observations  and calibration of
 the free parameters}

The power injected into the solar oscillations 
is related to the rms value $v_s$ of the surface velocity  as
\eqn{
v_s^2   = \xi_r^2(r_s)  \;  P \, / \, 2 \eta I
\label{eqn:vs2}
}
where  $r_s$ is the radius at which oscillations are measured.
Observations of the solar oscillations provide $v_s$ and $\eta$ such that 
$P(\omega)$ can be evaluated
 according to Eq.(\ref{eqn:vs2}). 

The power $P(\omega)$  is computed for  a calibrated solar model obtained with the CESAM code \citep{Morel97}. Convection is described according to the classical mixing-length theory \citep{Bohm58}. The oscillation properties were obtained from the adiabatic FILOU pulsation code of \citet{Tran95}.
The physical ingredients of the model are detailed in \citet{Samadi00}.

The free parameters are then  adjusted to obtain the best fit of the  frequency  dependence and the maximum amplitude   
to the solar observations by \citet{Libbrecht88}. 

\begin{figure}[ht]
\resizebox{\hsize}{!}{\includegraphics{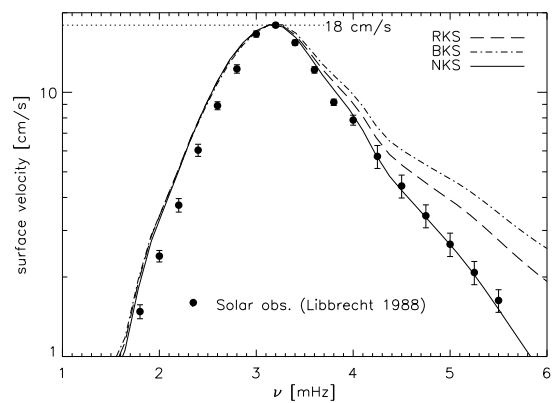}}
\caption{Computed surface velocity $v_s$ (Eq.\ref{eqn:vs2}) assuming the different turbulent spectra of Fig.(\ref{fig:spc_cinetique}). $\lambda$ and $\beta$ values result from fitting the computed $v_s$ to the solar seismic observations by \citet{Libbrecht88} using the observed damping rate $\eta$. }
\label{fig:VRSEP_cspc_fit}
\end{figure}

Results of the fitting are shown in Fig. \ref{fig:VRSEP_cspc_fit}.
All spectra of Fig \ref{fig:spc_cinetique} fit well the solar observations at
 low frequency ($\nu \lesssim 3.5$~mHz) 
while the main differences are observed at high frequency. The
overall best agreement is obtained with the NKS.
Details of the resulting adjustments are given in
 \citet{Samadi00II}.

\section{Scanning the HR diagram}

\subsection{The models}
We focus on low  intermediate mass stars ($1 \lesssim M \lesssim 2 M_\odot$) 
because the existence of  an outer convective  shell  
enables stochastic excitation. The thickness of the convective 
shell depends  in particular on 
the star luminosity ($L$) and effective temperature ($T_{\mathrm{eff}}$).
We thus consider six models well distributed in the solar-like oscillation region (Table \ref{tab:models_param}).
These models are based on the fundamental parameters of several bright stars which were selected as targets for EVRIS \citep{Baglin93}.
The  equilibrium models and the associated eigenfunctions of  
these stars are obtained in the same way as for the solar model.

\begin{table}[ht]
\begin{center}
\begin{tabular}{rrccccc}  
\multicolumn{2}{l}{Models}   & $L$ & $T_{\mathrm{eff}}$ &  $M$ & $ \nu_c$& Age\\
& & [$L_\odot$] & [K] & [$M_\odot$] & [mHz] & [Gyr]\\
\hline 
$A$ &$\alpha$~Tri & 12.1 & 6350 &    1.68 & 1.0 &  1.79 \\ 
$B$ &$\eta$~Boo &  9.0 & 6050  & 1.44 & 1.0 & 3.05 \\ 
$C$ &Procyon  &  6.6 & 6400  & 1.46 & 1.5 & 2.40 \\ 
$D$ &$\beta$~Hyrdi  & 3.7 & 5740 & 1.08 & 1.5 & 7.33\\ 
$E$ &$\beta$~Vir & 3.5  &   6120  & 1.25 & 2.3  &  4.10\\ 
$F$& $\pi^3$~Ori  & 2.6 & 6420 &  1.25  & 3.6 & 1.76 
\end{tabular}
\end{center}
\caption{Stellar Parameters of selected stars. $\nu_c$ is the cut-off frequency. For each model we have given the corresponding bright star from the EVRIS preliminary list of targets. 
}
\label{tab:models_param}
\end{table}

\subsection{Star to star comparison}

\begin{figure*}[ht]
 \resizebox{\hsize}{!}{\includegraphics{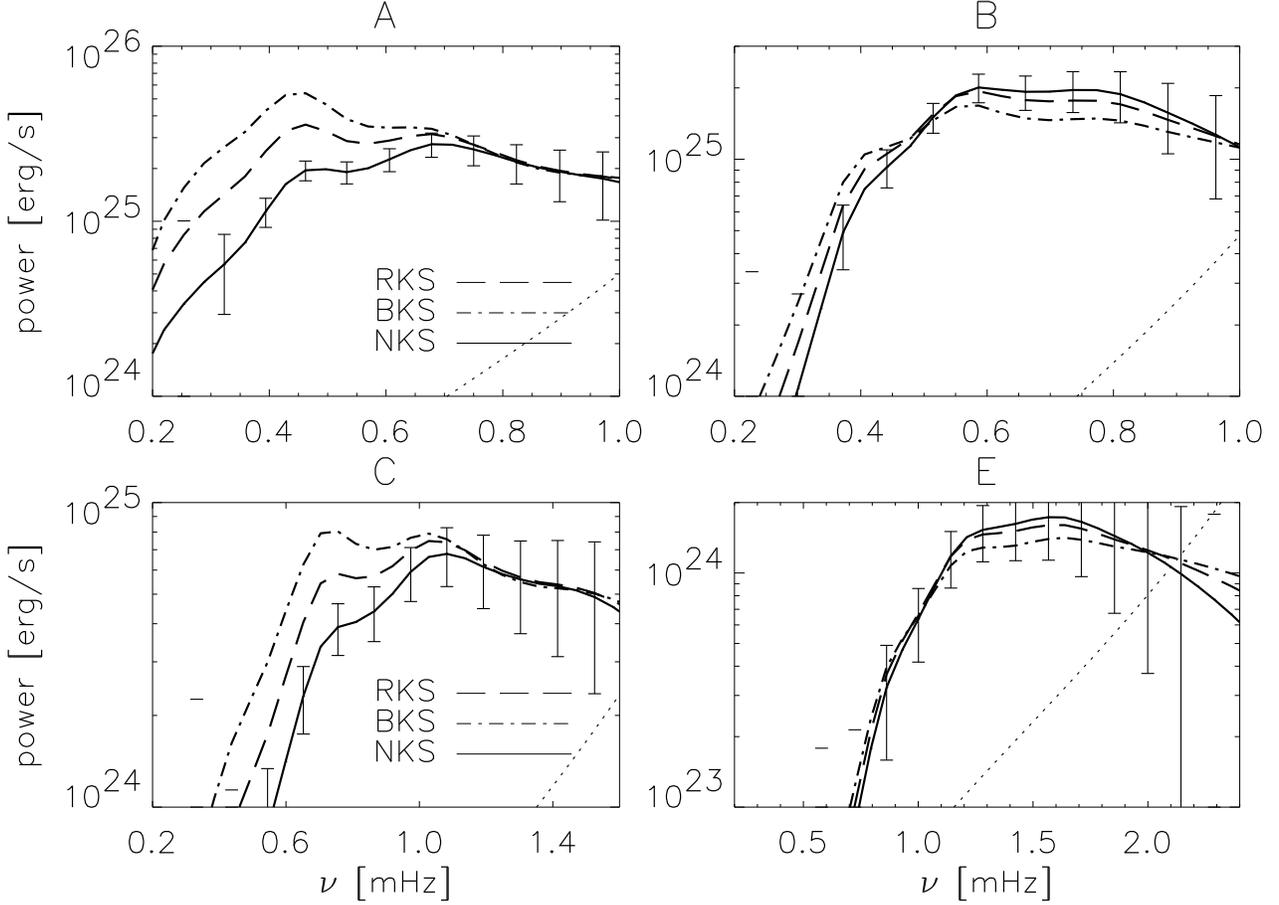}}
 \caption{Oscillation power $P$ versus frequency ($\nu=\omega/2\pi$)
 assuming the different spectra of Fig.1 for the stars~: $A$, $B$, $C$ and $E$.  
Vertical error bars $\Delta P$ (plotted on the solid line) were obtained from Eq.(\ref{eqn:DeltaP_P}) assuming
an  accuracy for $\eta/ 2\pi$ (resp. $\delta L/L$) of $0.1\,\mu$Hz (resp. $0.7$~ppm). 
For each star the detection threshold calculated according to
 Eq.(7) is represented by a dotted line.}
 \label{fig:pRSEP_stars2stars_spc}
\end{figure*}

\begin{figure*}[ht]
	\begin{center}
\resizebox{\hsize}{!}{\includegraphics{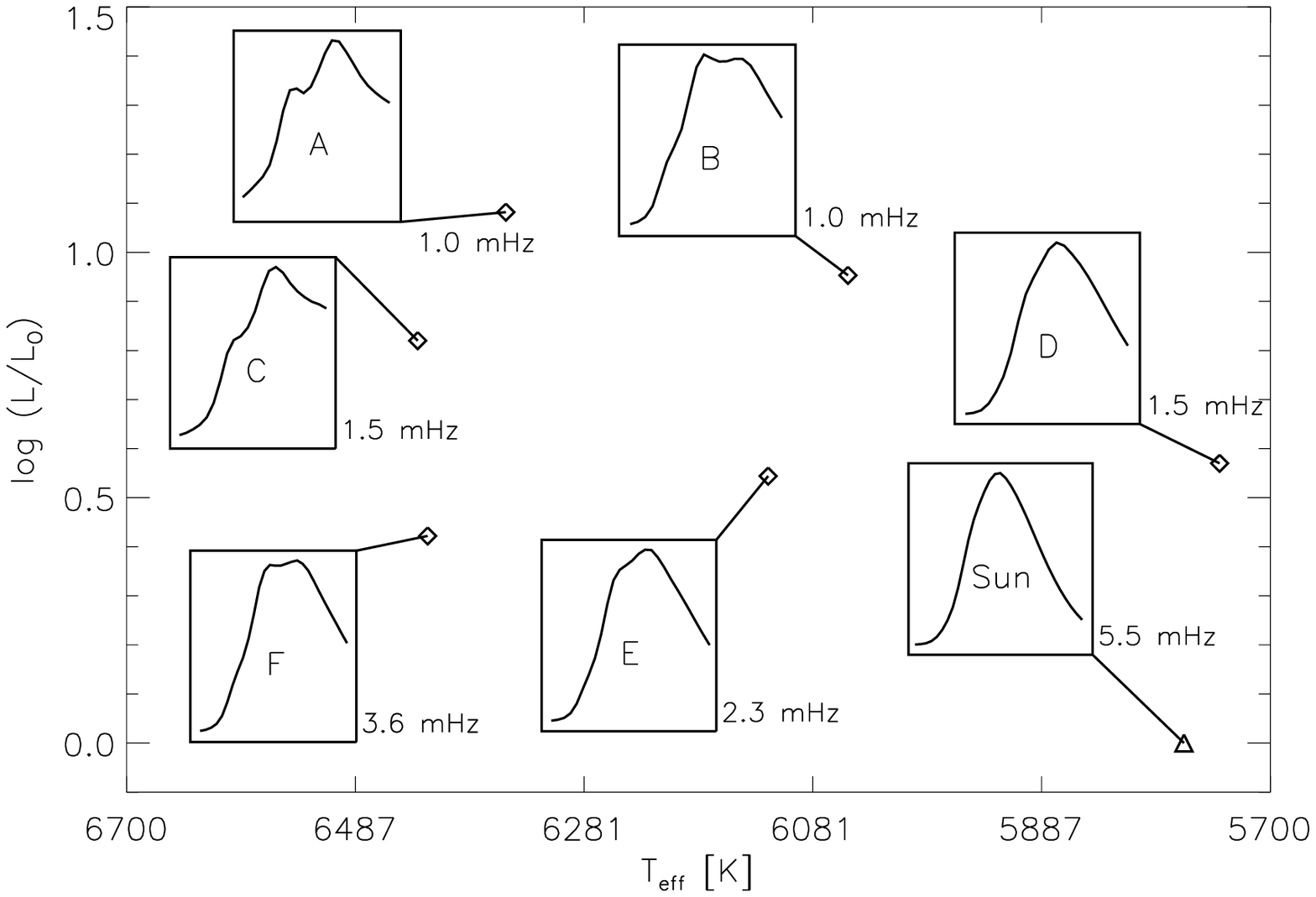}}
\end{center}
	\caption{Positions of the selected stars in the HR diagram (filled
squared). 
For each star the computed power spectrum assuming the NKS 
is shown in an associated box. Abscissa extend from 0 to the acoustic cut
off frequency : the frequency range of the oscillation power narrows
for more massive stars.}
	\label{fig:pRSEPnkcs_stars2stars_HR}
\end{figure*}

The acoustic power is computed for each  selected star 
and for each of  the three turbulent spectra of Fig.1 and is 
 plotted  for four of them  in Fig.\ref{fig:pRSEP_stars2stars_spc}.
Of our 5 stars, model $E$ is the closest to the Sun in the HR diagram
(Figure \ref{fig:pRSEPnkcs_stars2stars_HR}), hence its structure 
(assuming here the same chemical composition) is nearly the same than in the
solar case. Accordingly the  oscillation spectrum 
of  model $E$ is quite similar to the solar one  (except small differences
at high frequency) and therefore does not show
large differences when using different turbulent spectra (due to the solar
calibration mentioned in Sect. 2.2); only  small differences
can be seen at high frequency. Same conclusion can be drawn for model $D$.

For a hotter star, the oscillation spectra computed with different
turbulent spectra differ at low frequencies.
As it is illustrated with the error bars and the detection 
threshold in Fig.2, 
such differences are significant enough to constrain stellar 
turbulent spectra providing that accurate measurements 
of $P(\omega)$ are available.

Differences in oscillation power between model $E$ (equivalently the Sun) 
 and the other stars
 is related to the more extended outer  convective zone (CZ hereafter) 
of model $E$. 
As the thickness of its CZ is larger, excitation of 
 the low frequency oscillations 
involves all turbulent eddies.
Therefore as for the Sun the power injected 
into low frequency oscillations is mainly governed 
by the eigenfunction behavior ($\partial \xi_r / \partial r$ and I) 
and not by the behavior of the turbulence 
 in the energy injection range (i.e at large turbulent scale).

For the other hotter stars,  
the relative smaller size of their CZ causes
the excitation region to be localized in a thinner domain 
where changes in the properties of  $\xi_r$ are large.
At a given frequency the extension  of the excitation region
 changes with the considered turbulent spectra. 
Therefore the shape of $P(\omega)$ is very 
sensitive to both the properties of $\xi_r$ - thus to  
$\omega$ - and the behavior of the turbulent spectra. As a result, 
significantly large  differences in term of $P(\omega)$ are observed
when using different turbulent spectra.
This feature is also well explained for Procyon (model $C$) in \citet{Samadi00III}.

Besides a temperature effect, there is also a luminosity (or mass) effect
 although much smaller. This is illustrated with  model $F$ 
which is much hotter than model $E$ but 
shows only a very weak sensitivity to the choice of the turbulent spectrum 
when computing  $P(\omega)$ (not shown here). In that sense,  model $F$ is
 closer to the cooler model $E$ than to the  other stars with similar
 effective temperatures. It is an intermediate case as it is 
 less massive than  for instance  model $C$. 
Therefore the excitation region for this star extends deeper down 
compared to that of  Procyon (model $C$).



Figure \ref{fig:pRSEPnkcs_stars2stars_HR} presents a star to star comparison  
of the oscillation  power computed assuming the NKS.
The plot is performed in the HR diagram in order to depict 
the dependence of  the power spectrum $P(\omega)$
 with the effective temperature and the luminosity.
As the star becomes hotter and more luminous the frequency 
domain where $P(\omega)$  takes significant values becomes smaller. 
This is due to the decrease of the cut-off frequency with 
increasing values of  $T_{\mathrm{eff}}$ and $L$.


\subsection{Observational constraints}

Observations of solar-like oscillations in Procyon 
have yet  been only clearly  discovered by \citet{Martic99} and \citet{Barban99}
 in spectroscopic surface velocity measurements. However currently \citet{Butler00} claimed the detection of solar like oscillations in $\beta$~Hydri.

Most of the current ground based observations 
are mainly limited by the daily aliases. In particular 
they cannot yet provide the growth rates 
 $\eta$ which are necessary
in order  to compute $P(\omega)$ from Eq.(\ref{eq:P_omega_0}). 
Thus we need observations of solar-like oscillations performed in such a way 
as to avoid the daily aliases.
In particular space based experiments are particularly adapted for asteroseismology.
The forthcoming space project COROT (Baglin \& the Corot team, 1998)  
based on photometric measurements will reach a noise 
level of $0.7$ ppm \citep{Auvergne00} and will 
thus detect oscillation amplitudes comparable to the  solar ones  ($\sim 2$ ppm). 
Furthermore the instrument will continuously monitor 
several stars during $\sim 150$ days giving a 
frequency precision of $\sim 0.1 \,\mu$Hz such 
that accurate measurements of $\eta$ will be available.

Will this high accuracy 
enable us to constrain the theory of the stochastic excitation and the stellar turbulence ?
The answer depends on the  detection threshold and the accuracy in term of $P(\omega)$.
For our stars here, $\eta$ are not currently available.
It is  possible however to evaluate crudely a detection threshold for $P(\omega)$.
Indeed, the power is related to the root mean-square of the surface velocity $v_s$ according to Eq.(\ref{eqn:vs2}) where $r_s$ is set to the radius at which $T=T_{\mathrm{eff}}$.
In the adiabatic assumption $v_s$ is  in turn simply related to the luminosity fluctuation $\delta L$ approximatively as \citep{Kjeldsen95}
\eqn{		
\delta L/ L  \propto v_s \,  T_{\mathrm{eff}}^{-1/2}
\label{eqn:dl_vs}
}
Eq.(\ref{eqn:vs2}) and Eq.(\ref{eqn:dl_vs}) enable us to derive an approximative  relation
\eqn{
\frac{P}{ P_\odot } \approx \left ( \frac{\delta L /L}{  \delta L_\odot /L_\odot   } \right )^2 \, \frac{ T_\mathrm{eff} \, \eta \,  I \,  \xi_{r}^{-2} } { \left ( T_{\mathrm{eff}} \, \eta I \, \xi_{r}^{-2} \right )_\odot } 
\label{eqn:P_Psun}
}
where quantities with a  $\odot$ are relative to the Sun.
\citet{Goldreich91} derived a simplified equation for $\eta$
 which  allows us to express $\eta /\eta_\odot$ as \citep[see][section 3.4.4]{Houdek96}
\eqn{
\frac{\eta}{\eta_\odot} = \frac{ L \,  ( H \, c_s^{-2} \,  \omega \, R ) ^2 \, I^{-1} }{L_\odot \,  ( H \, c_s^{-2} \,  \omega \, R )_\odot ^2 \, I_\odot^{-1} }
\label{eqn:eta_etaSun}
}
where $c_s$ and $H$ are the sound speed and the pressure scale height respectively evaluated at the top of the convection zone and $R$ is the star radius.
Let  $s$ be the detection threshold in 
terms of relative luminosity fluctuation $\delta L/L$
 expressed in unity of $(\delta L / L)_\odot$.
 The detection threshold $P^{\rm th}(\omega)$ in terms of oscillation power
 can be expressed with the help of 
Eqs(\ref{eqn:P_Psun},\ref{eqn:eta_etaSun}) as 
\eqna{
P^{\rm th}(\omega) =  b \;  s^2  P_\odot^\mathrm{max} \frac{ R^2 \ \xi_{r}^{-2}(\omega)\; \, \omega^2 } {R_\odot^2 \,  \xi_{r,\odot }^{-2}(\omega_\mathrm{max}) \, \omega_\mathrm{max}^2 }  \\
 \mathrm{with} \; b  = T_\mathrm{eff} \, L \, H^2 \, c_s^{-4} \,  / \,  \left ( T_\mathrm{eff} \, L \, H^2 \, c_s^{-4}  \right )_\odot \nonumber
\label{eqn:Pth} 
}
where $\omega_{\rm max}$ and $P_\odot^\mathrm{max}$ are the 
frequency position and the power  at the maximum value 
of $\delta L_\odot/L_\odot$.
In addition from Eq.(\ref{eqn:eta_etaSun}) relative error bar in term of $P$ can be expressed as
\eqn{
\frac{\Delta P}{P} = 2 \, \frac{\Delta \delta L}{\delta L} + \frac{\Delta \eta}{\eta}
\label{eqn:DeltaP_P}
}
where $\eta$ and $\delta L$ are evaluated according to Eqs.(\ref{eqn:eta_etaSun},\ref{eqn:P_Psun}).

In Fig.\ref{fig:pRSEP_stars2stars_spc} the  
detection threshold $P^{th}(\omega)$ is plotted for the  case  $s=1$ which corresponds to 3 times the noise level (rms) of the COROT instrument. In addition the error bars $\Delta P$ according to Eq.(\ref{eqn:DeltaP_P}) is plotted assuming COROT's performances, {\it i.e.} $\Delta \eta / 2 \pi \sim 0.1 \, \mu$Hz and $\Delta \delta L \sim 0.7$~ppm. Note that, as for the Sun, $\Delta P$ is larger at high and low frequency because $\delta L$ becomes very smaller.

We conclude that the differences between the spectra for the hotter stars (models $A$ and $C$)  are sufficiently large compared to error bars to identify the best turbulent spectrum.

\section*{Conclusion}

This work shows that for  stars hotter and more massive than the Sun,
the formulation of the power $P(\omega)$ injected into the solar-like oscillations 
is very sensitive to the way the turbulent stellar spectrum is modeled.
For the Sun it is presently possible to constrain the solar 
turbulent spectrum from observations of the granulation. 
Such information is  not  directly available for other stars.
Evaluation of $P(\omega)$  derived from the seismic
 observations of the solar-like oscillations  (amplitudes and damping rates) 
  will therefore provide constraints on  the stellar turbulence.

In addition for a given spectrum it is found that
the shape of $P(\omega)$  versus $\omega$ can change quite significantly  from one
star to the other.
These differences are related to the size of the convection zone 
which depends on the fundamental parameters of the stars.
A star to star comparison of the power derived from  observations 
will thus provide additional constrains. 

The accuracy of an observational  determination of $P(\omega)$ 
depends on the accuracy of the measurements of the associated  oscillation fluctuations 
$\delta L$ and the oscillations damping rates $\eta$.
A crude relation derived in the present work allows to evaluate 
 the COROT detection threshold and the accuracy 
in terms of $P(\omega)$. 
It suggests  that futur space observations of such solar-like oscillating stars will measure acoustic powers $P(\omega)$  with an accuracy which will able us to identify the turbulent spectrum closest to reality. 


\begin{thebibliography}{24}
\expandafter\ifx\csname natexlab\endcsname\relax\def\natexlab#1{#1}\fi
\expandafter\ifx\csname url\endcsname\relax
  \def\url#1{{\tt #1}}\fi
\expandafter\ifx\csname urlprefix\endcsname\relax\def\urlprefix{URL }\fi

\bibitem[{{Auvergne} \& {the COROT Team}(2000)}]{Auvergne00}
{Auvergne} M., {the COROT Team}, april 2000, In: The Third MONS Workshop :
  Science Preparation and Target Selection, 135--138

\bibitem[{{Baglin} \& {The Corot Team}(1998)}]{Baglin98}
{Baglin} A., {The Corot Team}, 1998, In: IAU Symp. 185: New Eyes to See Inside
  the Sun and Stars, vol. 185, 301

\bibitem[{{Baglin} et~al.(1993){Baglin}, {Weiss}, \&
  {Bisnovatyi-Kogan}}]{Baglin93}
{Baglin} A., {Weiss} W., {Bisnovatyi-Kogan} G., 1993, In: {Baglin} A.,
  W.~{Weiss} W. (eds.) Inside the stars, 756

\bibitem[{{Balmforth}(1992)}]{Balmforth92c}
{Balmforth} N.J., Apr. 1992, \mnras, 255, 639

\bibitem[{{Barban} et~al.(1999){Barban}, {Michel}, {Martic} et~al.}]{Barban99}
{Barban} C., {Michel} E., {Martic} M., et~al., Oct. 1999, \aap, 350, 617

\bibitem[{{B\"ohm - Vitense}(1958)}]{Bohm58}
{B\"ohm - Vitense} E., 1958, Zeitschr. Astrophys., 46, 108

\bibitem[{{Butler} \& {et~al.}(2000)}]{Butler00}
{Butler} P., {et~al.}, 2000, (in preparation)

\bibitem[{{Espagnet} et~al.(1993){Espagnet}, {Muller}, {Roudier}, \&
  {Mein}}]{Espagnet93}
{Espagnet} O., {Muller} R., {Roudier} T., {Mein} N., Apr. 1993, \aap, 271, 589

\bibitem[{{Goldreich} \& {Keeley}(1977)}]{GK77}
{Goldreich} P., {Keeley} D.A., Feb. 1977, \apj, 212, 243

\bibitem[{{Goldreich} \& {Kumar}(1991)}]{Goldreich91}
{Goldreich} P., {Kumar} P., jun 1991, \apj, 374, 366

\bibitem[{{Goldreich} et~al.(1994){Goldreich}, {Murray}, \& {Kumar}}]{GMK94}
{Goldreich} P., {Murray} N., {Kumar} P., Mar. 1994, \apj, 424, 466

\bibitem[{{Houdek}(1996)}]{Houdek96}
{Houdek} G., 1996, Pulsation of Solar-type Stars, Ph.D. thesis, Institut f\"ur
  Astronomie , Wien

\bibitem[{{Kjeldsen} \& {Bedding}(1995)}]{Kjeldsen95}
{Kjeldsen} H., {Bedding} T.R., Jan. 1995, \aap, 293, 87

\bibitem[{{Libbrecht}(1988)}]{Libbrecht88}
{Libbrecht} K.G., Nov. 1988, \apj, 334, 510

\bibitem[{{Martic} et~al.(1999){Martic}, {Schmitt}, {Lebrun} et~al.}]{Martic99}
{Martic} M., {Schmitt} J., {Lebrun} J.C., et~al., 1999, \aap, 351, 993


\bibitem[{{Morel}(1997)}]{Morel97}
{Morel} P., Sep. 1997, \aaps, 124, 597

\bibitem[{{Musielak} et~al.(1994){Musielak}, {Rosner}, {Stein}, \&
  {Ulmschneider}}]{Musielak94}
{Musielak} Z.E., {Rosner} R., {Stein} R.F., {Ulmschneider} P., Mar. 1994, \apj,
  423, 474

\bibitem[{{Nesis} et~al.(1993){Nesis}, {Hanslmeier}, {Hammer} et~al.}]{Nesis93}
{Nesis} A., {Hanslmeier} A., {Hammer} R., et~al., Nov. 1993, \aap, 279, 599

\bibitem[{{Samadi} \& {Goupil}(2000)}]{Samadi00I}
{Samadi} R., {Goupil} M.J., 2000, \aap (submitted)

\bibitem[{{Samadi} \& {Houdek}(2000)}]{Samadi00}
{Samadi} R., {Houdek} G., april 2000, In: The Third MONS Workshop : Science
  Preparation and Target Selection, 27--32

\bibitem[{{Samadi} et~al.(2000{\natexlab{a}}){Samadi}, {Goupil}, \&
  {Lebreton}}]{Samadi00II}
{Samadi} R., {Goupil} M.J., {Lebreton} Y., 2000{\natexlab{a}}, \aap (submitted)

\bibitem[{{Samadi} et~al.(2000{\natexlab{b}}){Samadi}, {Houdek}, {Goupil}, \&
  {Lebreton}}]{Samadi00III}
{Samadi} R., {Houdek} G., {Goupil} M.J., {Lebreton} Y., 2000{\natexlab{b}}, in
  preparation

\bibitem[{{Tran Minh} \& {Leon}(1995)}]{Tran95}
{Tran Minh} F., {Leon} L., 1995, In: Physical Process in Astrophysics, 219

\end{thebibliography}



\end{document}